\documentclass[%
 aip,
 amsmath,amssymb,
 reprint,%
]{revtex4-1}

\usepackage{graphicx}% Include figure files
\usepackage{dcolumn}% Align table columns on decimal point
\usepackage{bm}% bold math
\usepackage[mathlines,pagewise]{lineno}% Enable numbering of text and display math
\usepackage[utf8]{inputenc}
\usepackage[T1]{fontenc}
\usepackage{mathptmx}
\usepackage{afterpage}
\usepackage{etoolbox}

\usepackage{tikz}
\usepackage{amssymb}
\usepackage{float} 

\DeclareMathOperator*{\argmin}{arg\,min}

\usepackage{diagbox, xcolor}
\newcommand{\tL}{\tilde{L}}
\newcommand{\bfx}{\mathbf{x}}
\newcommand{\bfxx}{\boldsymbol{x}}
\newcommand{\bfXX}{\boldsymbol{X}}

\newcommand{\bfX}{\mathbf{X}}
\newcommand{\bfa}{\mathbf{a}}
\newcommand{\bfW}{\mathbf{W}}
\newcommand{\bfb}{\mathbf{b}}
\newcommand{\bfTheta}{{\Theta}}
\makeatletter
\def\@email#1#2{%
 \endgroup
 \patchcmd{\titleblock@produce}
  {\frontmatter@RRAPformat}
  {\frontmatter@RRAPformat{\produce@RRAP{*#1\href{mailto:#2}{#2}}}\frontmatter@RRAPformat}
  {}{}
}%
\makeatother
\begin{document}

\preprint{AIP/123-QED}

\title{Physics-informed neural networks for gravity currents reconstruction from limited data}
% Force line breaks with \\
\author{Mickaël Delcey}
 \affiliation{LEMTA, Université de Lorraine, CNRS, 2, Avenue de la Forêt de Haye, B.P. 160, 54504 Vandœuvre-lès-Nancy, France}%Lines break automatically or can be forced with \\
\author{Yoann Cheny}%
\email[email address: ]{yoann.cheny@univ-lorraine.fr}
\affiliation{LEMTA, Université de Lorraine, CNRS, 2, Avenue de la Forêt de Haye, B.P. 160, 54504 Vandœuvre-lès-Nancy, France}
% \email{yoann.cheny@univ-lorraine.fr}

\author{Sébastien Kiesgen de Richter}
\affiliation{LEMTA, Université de Lorraine, CNRS, 2, Avenue de la Forêt de Haye, B.P. 160, 54504 Vandœuvre-lès-Nancy, France}
\affiliation{Institut Universitaire de France (IUF)}
\date{\today}% It is always \today, today,

\begin{abstract}

The present work investigates the use of physics-informed neural networks (PINNs) for the three-dimensional (3D) reconstruction of unsteady gravity currents from limited data. In the PINN context, the  flow fields are reconstructed by training a neural network whose objective function penalizes the mismatch between the network predictions and the observed data and embeds the underlying equations using automatic differentiation. 
This study relies on a high-fidelity numerical experiment of the canonical lock-exchange configuration. This allows us to benchmark quantitatively the PINNs reconstruction capabilities on several training databases that mimic state-of-the-art experimental measurement techniques for density and velocity. Notably, spatially averaged density measurements by light attenuation technique (LAT) are employed for the training procedure. We propose an experimental setup that combines density measurement by LAT and two independent planar velocity measurements by particle image velocimetry (PIV). The so-called LAT-2PIV setup gives the most promising results for flow reconstruction by PINNs, with respect to its accuracy and cost efficiency.
\end{abstract}

\maketitle

\section{Introduction}\label{intro}
Recovering fully three-dimensional (3D) fluid flow from limited, often planar, observations of the velocity, temperature or density is a challenging problem that can find many applications in science and engineering fields. Machine learning enabled recent progress in this direction\cite{callaham2019robust}, but 3D flow reconstruction remains out of the scope of purely data-driven approaches : interpolative techniques are likely to fit the observational data very well, but are equally likely to fail to generalize beyond observations. To address this issue observational data should be combined with the partial derivative equations (PDE) governing the system under observation. Variational data assimilation (VDA) techniques~\cite{mons2016reconstruction} follow this idea by fueling computational fluid dynamics (CFD) solvers with flow observations. While leveraged on a rigorous mathematical framework, VDA is however a very intensive CPU task : each step of the resulting iterative procedure requires direct run of a CFD solver and estimation of its adjoint. For this reason the use of VDA to 3D flow reconstruction is very limited~\cite{chandramouli20204d}.

The family of physics-informed neural networks~\cite{raissi2019physics} (PINNs) opened a door for flow reconstruction. This new class of deep learning methods seamlessly integrates the observation data with the PDEs and has been successfully applied for solving numerous forward and inverse problems in fluid mechanics~\cite{cai2022physics}, solid mechanics~\cite{haghighat2021physics}, heat transfer~\cite{cai2021physics}, see Ref.~\onlinecite{cuomo2022scientific} for an exhaustive review. The reasons of the success of PINNs are twofold : the problem solution is approximated by artificial neural networks (ANNs) that are endowed with unlimited expressivity~\cite{hornik1989multilayer} and time and space operators in PDEs are efficiently computed at machine precision with automatic differentiation (AD). 

In Ref.~\onlinecite{HFM}, PINNs have been applied for reconstructing the 3D flow past a circular cylinder. The method allows to build surrogates for velocity-pressure fields in a region of interest where volumetric data on the concentration of a passive scalar are given. The resulting approach has many seducing features that have been assessed on synthetic data : (i) the region of interest where data are provided can be chosen arbitrarily since the algorithm is agnostic of boundary conditions and geometry, (ii) the method is robust to low resolution and noise and (iii) the surrogates models are fully differentiable which allows computation of any quantity of interest (stresses, vorticity, etc...). An extension of this study~\cite{cai2022physics} has shown that PINNs allow accurate reconstruction of the flow in the case of limited data. However, up to five planar velocity measurements were used for training the PINN, which would be experimentally prohibitive. Later, the method has been applied for the first time on real experimental data~\cite{espresso}. The studied case is natural convection over an espresso cup for which detailed volumetric measurements of the unsteady temperature field are given by mean of Tomographic Background Oriented Schlieren. The inferred velocity fields have been qualitatively validated with independent Particle Image Velocimetry (PIV) measurements, which has revealed the invaluable potential of PINNs for experimental Fluid Mechanics.

In this work we apply for the first time the PINN method to gravity currents, which are an ubiquitous class of geophysical flows induced by density gradient \cite{ungarish2009introduction}. 
%Gaining in-depth understanding of gravity currents requires a detailed 3D picture of both density, velocity and pressure fields. However, most of nowadays experimental techniques allow planar or averaged measurements of density that are often combined with planar PIV observations, leading experimentalists to rely on questionable assumption of flow homogeneity in one direction.  This motivates the use of PINNs for reconstructing flow information in the discarded direction.
These flows are most often 3D, transient, unsteady and inhomogeneous making the realization of experimental measurements resolved both spatially and temporally very delicate and often incomplete. The objective of this paper is to show that the PINNs approach can infer with high accuracy the hydrodynamic fields of this type of flows from incomplete data (two-dimensional (2D) or integrated measurements). 

In this work we focus on a numerical experiment performed with the spectral solver NEK5000 of the Lock-exchange flow configuration which is widely studied at the laboratory scale. The high fidelity numerical solution is subsequently used for designing several training data sets that mimic state-of-the art experimental setups and serves as ground truth for validation purpose which is described in Section \ref{prob}. In Section \ref{methodo} we present the extension of the PINN approach to gravity currents and propose a modified loss function that takes into account spatially averaged data. The resulting approach is applied in section \ref{results} to three existing experimental setups, for which the accuracy of the reconstructed fields is discussed. We also investigate the design an "optimal" experimental setup with respect to two criteria : the complexity/price of the setup and the accuracy of the inferred fields. Concluding remarks are given in Section \ref{conclusion}.

\section{\label{prob} Synthetic data for the lock-exchange problem}
This section describes the canonical lock-exchange problem and how the governing equations have been numerically solved with the code Nek5000 \cite{nek5000-web-page}. From the numerical solution, four datasets were carefully designed to mimic state-of-the-art measurement techniques, which is presented in details.

\noindent The lock-exchange flow configuration (see Fig.~\ref{fig:cuve}) has long served as a paradigm configuration for studying the spatio-temporal evolution of gravity currents, it consists of a 3D rectangular channel of size $\tL_x\times \tL_y\times \tL_z$  (a tilde denotes a dimensional quantity here) that is initially filled with two miscible fluids separated by a membrane located at $\tilde{x}=\tilde{l}_x$. The fluids in the left- and right-hand compartments have the respective densities $\tilde{\rho}_1$ and $\tilde{\rho}_2$, with $\tilde{\rho}_1 > \tilde{\rho}_2$, the density gradient is caused for instance by salinity difference. Once the membrane is withdrawn, a gravity current containing the heavier fluid starts propagating rightwards along the bottom wall. 

\subsection{Governing equations}
In order to render the equations dimensionless~\cite{hartel2000analysis}, we use $\tL_y$ as the characteristic
length scale and the characteristic velocity is the buoyancy velocity defined as~:
\begin{equation}\label{eq:buyvel}
    \tilde{u}_b=\sqrt{g\frac{\tilde{\rho_1}-\tilde{\rho_2}}{\tilde{\rho_a}}\tilde{L}_y}, \quad \text{with}\quad\tilde{\rho}_a=\frac{\tilde{\rho_1}+\tilde{\rho_2}}{2}.
\end{equation}
We consider small density difference $\tilde{\rho}_1 - \tilde{\rho}_2\ll \tilde{\rho}_1$ for which the Boussinesq approximation can be adopted, thus the governing equations read in dimensionless form~:

% We demonstrate the validity of the model with ideal numerical data generated by CFD for the case of gravity currents produced by lock-exchange, with Re = 5500. We use the spectral element code Nek5000 over the 3D dimensionless spatial and time domain $\Omega = [0,6.0] \times [0,1.0] \times [0,0.4]$ and $T = [0,9.7]$. We assume symmetry condition at $z=0.1$, boundary conditions at the bottom $z=0$, down  $y = 0$, left and right $x=0$ and $x=6.0$. The governing set of equations on this case are the dimensionless incompressible Navier-stokes equations and the mass transport equation expressed below :
% In dimensionless form the governing equations thus read :
\begin{subequations}
\label{eq:NS}
\begin{align}
    & \dfrac{\partial u}{\partial t} + (\boldsymbol{v} \cdot \nabla )u  +\dfrac{\partial p}{\partial x} - \dfrac{1}{Re}\nabla ^{2} u  \label{eq:momx}=0\\
    & \dfrac{\partial v}{\partial t} + (\boldsymbol{v} \cdot \nabla )v +\dfrac{\partial p}{\partial y} - \dfrac{1}{Re}\nabla ^{2} v +\rho =0\label{eq:momy}\\
    & \dfrac{\partial w}{\partial t} + (\boldsymbol{v} \cdot \nabla )w +\dfrac{\partial p}{\partial z} - \dfrac{1}{Re}\nabla ^{2} w \label{eq:momz}=0\\
    & \dfrac{\partial \rho}{\partial t}+(\boldsymbol{ v} \cdot \nabla )\rho - \dfrac{1}{Re\,Sc}\nabla^{2} \rho \label{eq:mass}=0\\
    & \nabla \cdot \boldsymbol{ v} = 0\label{eq:cont},  
\end{align}
\end{subequations}
where  $\boldsymbol{v} = (u,v,w)^\intercal$ denotes the dimensionless velocity vector. The non-dimensional pressure $p$ and density $\rho$ are given by~:
\begin{equation}\label{eq:prhoadim}
    p=\frac{\tilde{p}}{\tilde{\rho}_a\tilde{u}_b^2},\quad \rho=\frac{\tilde{\rho}-\tilde{\rho}_2}{\tilde{\rho_1}-\tilde{\rho_2}}.
\end{equation}
The Reynolds number $Re$ and the Schmidt number $Sc$ arising in the dimensionless equations \eqref{eq:NS} are defined by~:
\begin{equation}\label{eq:ReSc}
    Re=\frac{\tilde{u}_b\tL_y}{\tilde{\nu}},\quad Sc=\frac{\tilde{\nu}}{\tilde{\kappa}},
\end{equation}
where $\tilde{\nu}$ is the kinematic viscosity and $\tilde{\kappa}$ denotes the molecular diffusivity of the chemical specie producing the density difference. 
\subsection{\label{cfd} Numerical solution}

An accurate representation of the interface between the miscible fluids is crucial in gravity currents simulations, which requires high-order numerical methods to compute steep gradients in the vicinity of the interface. To this end, the governing equations~\eqref{eq:NS} are solved with the code Nek5000 \cite{nek5000-web-page} that has been successfully employed in numerical investigations of gravity currents~\cite{ozgokmen2004three,ozgokmen2006product,ozgokmen2009reynolds}. The discretization scheme in Nek5000 is based on the spectral element method~\cite{patera1984spectral} with exponential convergence in space and $3^\text{rd}$-order timestepping scheme. 

We consider the dimensionless computational domain $\Omega = [0,6.0] \times [0,1.0] \times [0,0.2]$ for $t\in[0,10]$ with $Re=5500$, which corresponds to the experimental conditions investigated in a companion article. Whilst for liquids such as salt water $Sc\simeq 700$, we consider here $Sc=1$. This assumption is commonly made to reduce computational costs and has low influence on the gravity current front~\cite{marshall2021effect}.
\noindent A free-slip condition (no normal flow) is applied at $y=0.4$, a symmetry condition is imposed at $z=0.2$ and no slip-conditions are applied on the remaining domain boundaries. Grid independent results were obtain on a computational mesh of $90\times45\times19$ elements where unknowns are represented by $7^\text{th}$-order Lagrange interpolating polynomials, and with a fixed timestep $\Delta t = 2.10^{-2}$ giving a Courant–Friedrichs–Lewy number less than $0.5$.

\definecolor{op}{gray}{0.9}
\begin{figure}[H]
    \centering 
     \includegraphics[scale = 0.58]{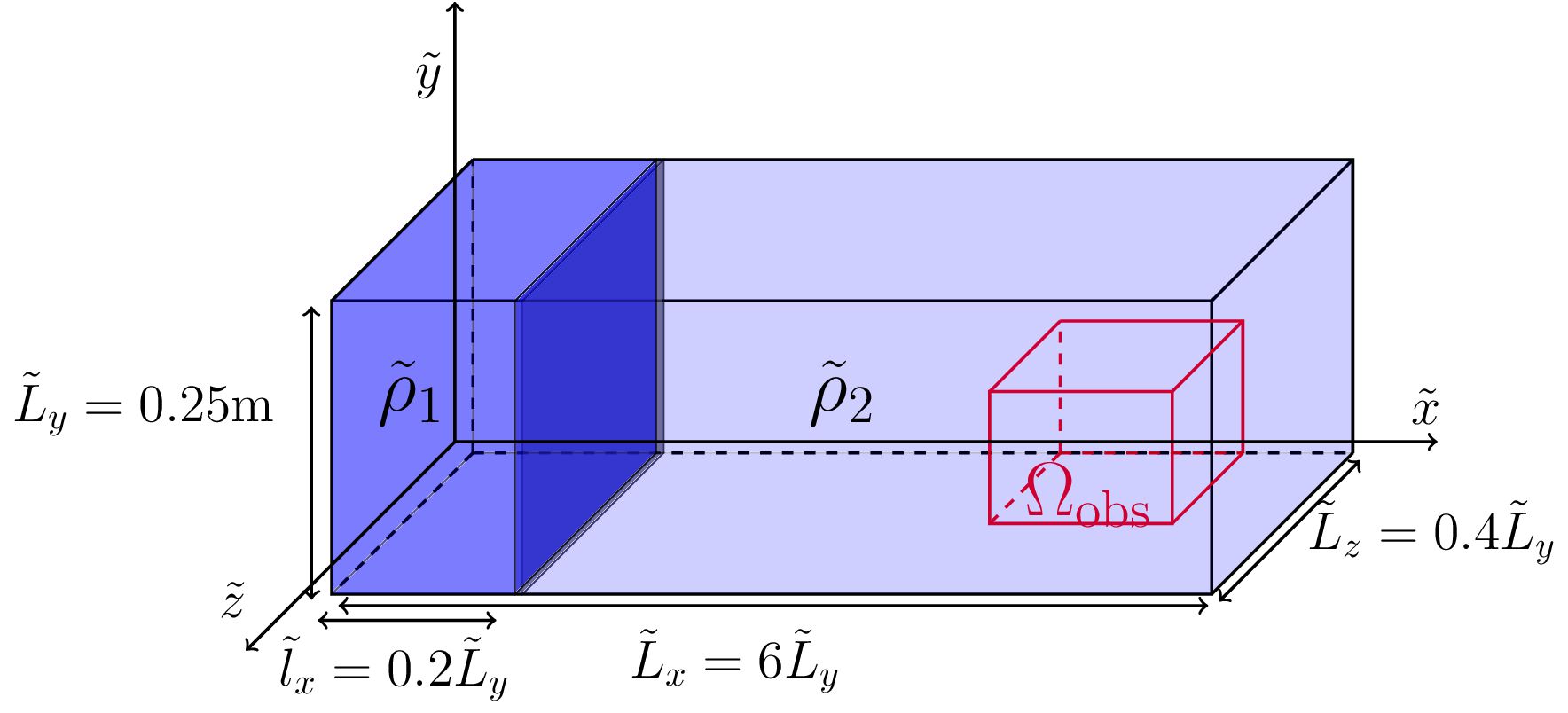}
    \caption{Sketch of the lock-exchange configuration. Initially two fluids of densities $\tilde{\rho}_1$ and $\tilde{\rho}_2$ are separated by a removable membrane located at $\tilde{x}=\tilde{l}_x$. The rectangular parallelepiped with red edges is the observational region of interest $\Omega_{\text{obs}}$.}
    \label{fig:cuve}
\end{figure}

% for the training, we concentrate on a spatial sub-domain of $\Omega$ : $\Omega_{1} = [3.5,5.0] \times [0.0,0.4] \times [0.0,0.4]$ and temporal sub-domain of $T$ : $T_{1}= [9.2,10.0]$

% \begin{figure}[H]
%     \centering 
%     \includegraphics[scale = 0.6]{../images/plot3D_2.png}
%     \caption{\textbf{Illustration of the Computational domain.} The flow is simulated with the CFD code Nek5000, density field is shown on the domain. A smaller box is selected for the training.  }
%     \label{fig:my_label}
% \end{figure}
\subsection{\label{sec:testcase} Data generation}

\begin{figure}[H]
    \centering 
    \includegraphics[scale = 0.25]{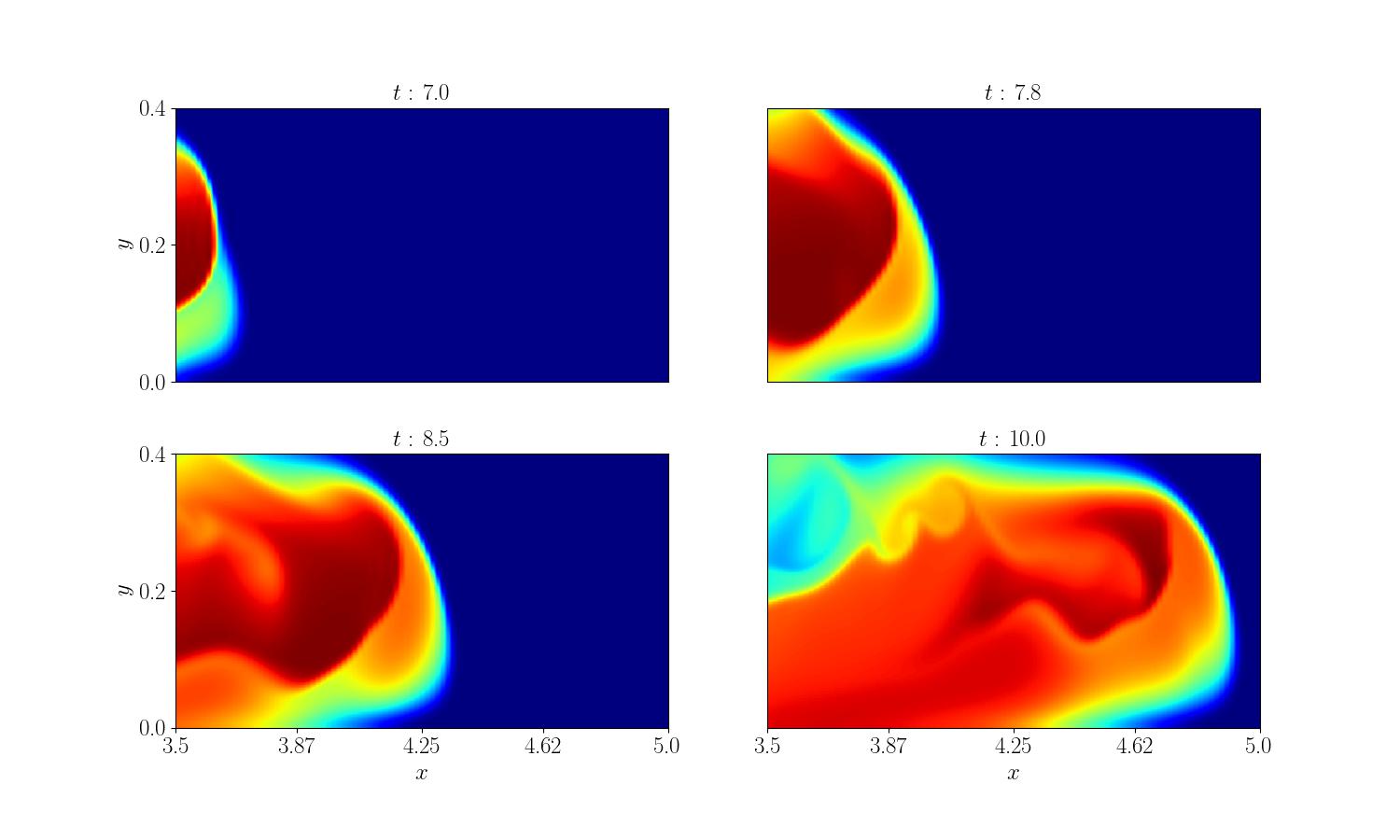}
    \caption{Time evolution of the density at $z = L_{z}/2$. The data acquisition starts at $t_\text{s}=7.0$ as the gravity current enters $\Omega_{\text{obs}}$ and until the front of the gravity current leaves $\Omega_{\text{obs}}$ at $t_\text{s}=10$. }
    \label{fig:snaptime}
\end{figure}

The reference dataset $\mathcal{D}^{\text{ref}}$ that will be used for PINN training is collected in the sub-domain $\Omega_{\text{obs}} = [3.5,5.0] \times [0.0,0.4] \times [0.0,0.2]\subset\Omega$ represented in red in Fig.~\ref{fig:cuve}, that is discretized with the uniform Cartesian grid defined by~:
\begin{equation}\label{eq:cartgrid}
   \!\!\mathcal{X}\times\mathcal{Y}\times\mathcal{Z}\equiv\biggl\{ x_i\biggr\}_{1\leq i\leq N_x}\!\!\!\times \biggl\{ y_j\biggr\}_{1\leq j\leq N_y}\!\!\!\times \biggl\{ z_k\biggr\}_{1\leq k\leq N_z},
\end{equation}
where $N_x=76$, $N_y=51$ and $N_z=18$. The data acquisition starts at $t_\text{s}=7.0$ as the gravity current enters $\Omega_{\text{obs}}$ and until the front of the gravity current leaves $\Omega_{\text{obs}}$ at $t_\text{s}=10$ (see Fig.~\ref{fig:snaptime}). We consider $N_t=50$ snapshots of the solution that are equally distributed in the time interval $T_{\text{obs}}= [7.0,10.0]$, which defines the discrete temporal domain~: 
\begin{equation}\label{eq:timemesh}
    \mathcal{T} =\biggl\{ t_n\biggr\}_{1\leq n\leq N_t}.
\end{equation}
Accordingly, the reference dataset $\mathcal{D}^{\text{ref}}$ contains more than 17M elements and is defined as~:
\begin{equation}\label{eq:refdata}
    \mathcal{D}^{\text{ref}} =\biggl\{\rho_{\bfx},u_{\bfx},v_{\bfx},w_{\bfx},p_{\bfx}\biggr\}_{\bfx\in\mathcal{M}},
\end{equation}
where $\rho_{\bfx}$ denotes the numerical solution computed at the element $\bfx$ of the spatio-temporal mesh $\mathcal{M}=\mathcal{T}\times\mathcal{X}\times\mathcal{Y}\times\mathcal{Z}$. We introduce an additional notation for discrete planes e.g., in the following $\mathcal{P}^z_{\!\!N_z}$ refers to the discretization of the symmetry plane $z=L_z/2$~:
\begin{equation}\label{eq:discplane}
    \mathcal{P}^z_{\!\!N_z} =\mathcal{X}\times\mathcal{Y}\times\biggl\{ z_{N_{z}}=\tfrac{L_z}{2}\biggr\}.
\end{equation}
From the reference dataset $\mathcal{D}^{\text{ref}}$, three measurement techniques frequently used to study gravity currents are considered~:

\noindent \textbf{[PLIF-PIV Case]} : This can be seen as the state-of-the-art method to study gravity currents \cite{balasubramanian2018entrainment,perez2018piv,garcia2020zonation,akbari2020injection}, where the density  and $(u,v)$ velocity components are observed in the symmetry plane. In a Planar Laser Induced Fluorescence (PLIF) system, a fluorescent dye is excited in a plane by a paired laser system. The resulting fluorescence is measured with a CCD camera and used to infer the corresponding density of the fluid. The two components of the velocity are experimentally measured with two-dimensional two components (2D-2C) Particle Image Velocimetry (PIV), which essentially requires a supplementary laser system. The training dataset $\mathcal{D}^{\text{PLIF-PIV}}$ that is used to mimic this experimental setup is defined as~:
\begin{equation}\label{eq:plifdata}
    \mathcal{D}^{\text{PLIF-PIV}} =\biggl\{\rho_{\bfx},u_{\bfx},v_{\bfx}\biggr\}_{\bfx\in\bfX^{\text{PLIF-PIV}}},\;\text{with}\;\bfX^{\text{PLIF-PIV}}= \mathcal{T}\times\mathcal{P}^{\,z}_{\!\!N_z}
\end{equation}

\noindent \textbf{[3D-LIF Case]} : The 3D-LIF technique \cite{partridge2019versatile} extends the principle of PLIF to 3D. The implementation of 3D-LIF is experimentally challenging and the volumetric measurements are achieved by mean of complex scanning mirror systems. In terms of implementation 3D-LIF is the most complex experimental device considered in this work and provides the largest training dataset $\mathcal{D}^{\text{3D-PIV}}$ that is expressed as~:
\begin{equation}\label{eq:3Dlifdata}
    \mathcal{D}^{\text{3D-LIF}} =\biggl\{\rho_{\bfx}\biggr\}_{\bfx\in \bfX^{\text{3D-LIF}}},\;\text{with}\;\bfX^{\text{3D-LIF}}= \mathcal{M}.
\end{equation}

\noindent \textbf{[LAT Case]} : The Light Attenuation Technique \cite{holford1996measurements} allows instantaneous measurements of the spanwise averaged density $\overline{\rho} = \frac{1}{L_z}\int_0^{L_z}\rho \,\text{d}z$. The averaging is performed along the optical path between a light source e.g., a DEL panel and a CCD camera. Whilst LAT is not widespread in the literature, it is a very accurate and yet easy to implement technique that is used notably in our research team \cite{dossmann2020asymmetric}. The corresponding dataset $\mathcal{D}^{\text{LAT}}$ reads~: 
\begin{equation}\label{eq:LATdata}
    \mathcal{D}^{\text{LAT}} =\biggl\{\overline{\rho}_{\bfxx}\biggr\}_{\bfX^{\text{LAT}}},\;\text{with}\;\bfX^{\text{LAT}}= \mathcal{T}\times\mathcal{X}\times\mathcal{Y}.
\end{equation}
The computation of $\overline{\rho}_{\bfxx}$ relies on the second-order accurate trapezoidal quadrature rule. This is the first time such averaged data are employed for PINN training.

\noindent We also consider two visualization techniques that have not yet been implemented experimentally. They are obtained by enriching the LAT case with additional (2D-2C) PIV measurements~:

\noindent\textbf{[LAT-PIV Case]} : This extends the LAT case with observations of the velocity components $(u,v)$ on the plane $z = \dfrac{L_{z}}{2}$. The corresponding dataset set reads :     
\begin{equation}\label{eq:LATPIVdata}
    \mathcal{D}^{\text{LAT-PIV}} = \mathcal{D}^{\text{LAT}}\cup\biggl\{u_{\bfx},v_{\bfx}\biggr\}_{\bfx\in \mathcal{T}\times\mathcal{P}^{\,z}_{N_z}}.
\end{equation}

\noindent\textbf{[LAT-2PIV Case]} : This extends the LAT-PIV case with observations of the velocity components $(u,v)$ on the plane $z = \dfrac{L_{z}}{4}$, which leads to the dataset~:
\begin{equation}\label{eq:LAT2PIVdata}
    \mathcal{D}^{\text{LAT-2PIV}} = \mathcal{D}^{\text{LAT-PIV}}\cup\biggl\{u_{\bfx},v_{\bfx}\biggr\}_{\bfx\in \mathcal{T}\times\mathcal{P}^{\,z}_{N_z/2}}.
\end{equation}
The salient properties and the experimental complexity of each case are summarized in Table~\ref{tab:data_prop}. We stress out that the implementation of PLIF-PIV system is experimentally more challenging than LAT-PIV or LAT-2PIV system. This is essentially due to the pairing between the fluorescent dye and the laser system, which makes the calibration of the experimental device significantly more complex.

\begin{table*}\caption{\label{tab:data_prop}Salient properties of the training datasets. The experimental complexity is evaluated by taking into account the required equipment and its calibration. The number of elements corresponds to the cardinal of the dataset i.e., $|\mathcal{D}^{\text{PLIF-PIV}}|=3\times50\times76\times51$. }
\begin{tabular}{| l|c|c|c | }
\hline
Case & Observed data   & Number of elements  & Estimation of experimental complexity  \\
\hline

\textbf{PLIF-PIV}    & $\rho,~ u, ~  v $  & $581\,400$ & $\star\star \star$ \\
\hline
\textbf{3D-LIF}    & $\rho$  & $3\,488\,400$&$\star\star\star\star \star$\\
\hline
\textbf{LAT}    & $\overline{\rho}$     &  $193\,800$ &  $\star$  \\
\hline
\textbf{LAT-PIV}   & $\overline{\rho},~ u, v $  & $581\,400$  & $\star \star$ \\
\hline
\textbf{LAT-2PIV}   & $\overline{\rho},~ u, ~ v $  & $969\,000$  & $\star\star \star$ \\
\hline
\end{tabular}
\end{table*}

\section{\label{methodo}Physics Informed Neural Networks for flow reconstruction}
\begin{figure*}
    \centering 
    \includegraphics[scale = 0.85]{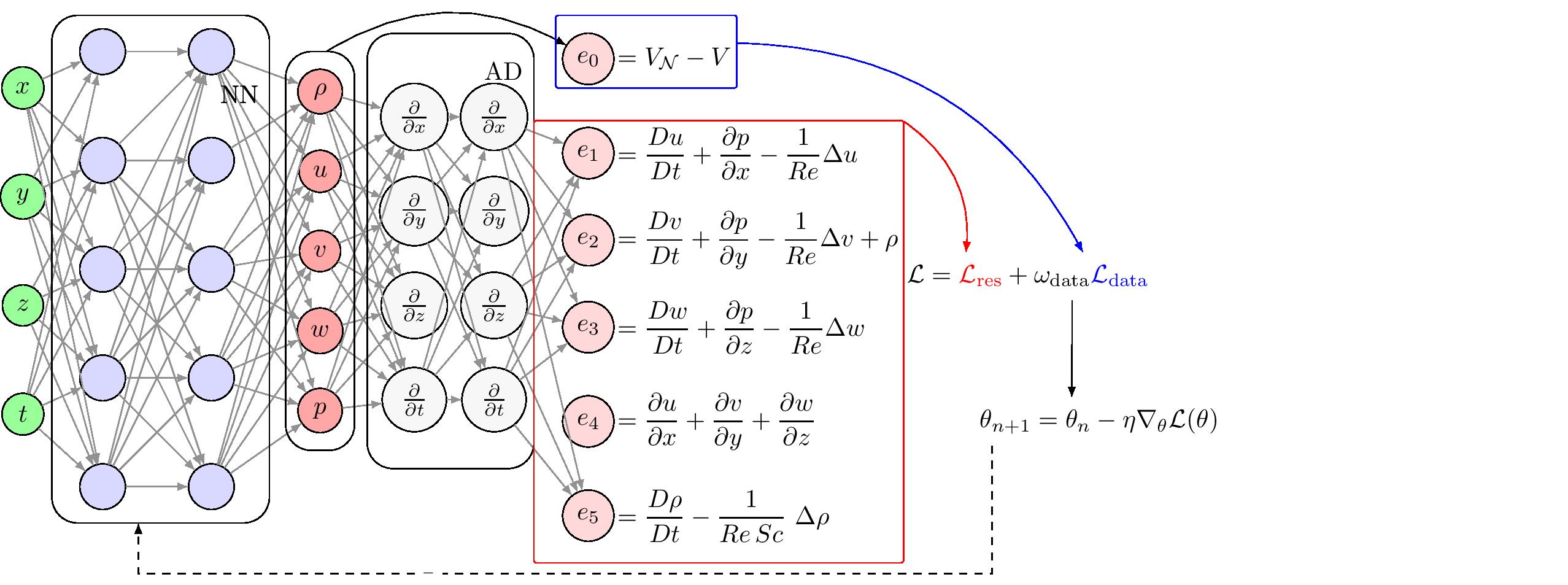}
    \caption{Physic informed neural network structure : a fully connected neural network take as input $\textrm{\textbf{x}}=(t,x,y,z) \in \mathbb{R}^{4}$ and predicts $\boldsymbol{\mathcal{N}}(\textrm{\textbf{x}}) = (\rho_{\mathcal{N}},u_{\mathcal{N}},v_{\mathcal{N}},w_{\mathcal{N}},p_{\mathcal{N}})$. The residuals of the governing equations $e_{1,4}$  are computed by automatic differentiation and $e_0$ denotes the mismatch between the observational data $V$ and the predictions $V_{\mathcal{N}}$, which are combined in the loss function $\mathcal{L}$. Finally, the weights and biases of the network are updated iteratively with gradient descent.  }
    \label{fig:sketchPINN}
\end{figure*}

The class of physics-informed machine learning methods regroups the different attempts for integrating physical information in the traditional deep learning workflow.  According to the physics-informed machine learning taxonomy of Kim etal.~\cite{kim2021knowledge}, PINNs correspond to a "ANN-differential equation-regularizing" pipeline~: an ANN is used as a surrogate model for the flow fields and the governing equations~\eqref{eq:NS} are leveraged to design an enriched loss function (Fig.~\ref{fig:sketchPINN}) that is minimized during training process, which is described in this section.
\definecolor{op}{gray}{0.9}

\subsection{\label{NN}Artificial neural networks}

The starting point of PINNs is to approximate the flow fields over the time space domain with a trained neural network (NN) $\mathcal{N}$, which reads~:
\begin{equation}\label{eq:res}
    (\rho_{\mathcal{N}},u_{\mathcal{N}},v_{\mathcal{N}},w_{\mathcal{N}},p_{\mathcal{N}})=\mathcal{N}(\bfx),
\end{equation}
with $\textrm{\textbf{x}}=(t,x,y,z) \in  T_{\text{obs}}  \times \Omega_{\text{obs}}$. A neural network can be expressed as a composition of a series of simple non-linear functions~:
\begin{equation}\label{eq:comp}
\mathcal{N}(\bfx) = f^{(L)}\circ \dots f^{(2)}\circ f^{(1)}(\bfx),
\end{equation}
where $L$ denotes the number of layers. The $l$-th layer contains $n_{l}$ neurons and, letting $\bfa^{(0)}=\bfx$, its output $\bfa^{(l)}\in\mathbb{R}^{n_{l}}$ can be recursively computed as~:
\begin{equation} \label{eq:layer}
 \bfa^{(l)} = f^{(l)}(\bfa^{(l-1)})= \sigma^{(l)}  \left( \bfW^{(l)} \bfa^{(l-1)} +\bfb^{(l)}  \right),
\end{equation} 
where $\sigma^{(l)}$ is a given elementwise non-linear activation function and $\bfW^{(l)}$ and $\bfb^{(l)}$ denote respectively the weight matrix and bias vector of the $l$-th layer. The trainable parameters of $\mathcal{N}$  are denoted by $\bfTheta= \{ \bfW^{(l)}, \bfb^{(l)}  \}_{1\leq l\leq L}$.  \\

\subsection{\label{sec:pinnloss}PINNs objective function}

Training the neural networks consists in searching the optimal set of parameters $\bfTheta^{*}$ that minimizes an objective function $\mathcal{L}$, which can be expressed as~:
\begin{equation} \label{eq:optprob}
\bfTheta^{*}= \underset{\bfTheta}{\argmin}  ~\mathcal{L}.
\end{equation}
Designing the objective function $\mathcal{L}$ is of the utmost importance since it characterises the quality of the solution. As an illustrative example, we consider the \emph{PLIF-PIV case} for which the dataset $\mathcal{D}^{\text{PLIF-PIV}}$ is provided. By adopting a naive data-driven approach, the objective function would correspond to the mismatch $\mathcal{L}_\text{data}$ between the NN predictions and the observed data $\mathcal{D}^{\text{PLIF-PIV}}$, namely~:
\begin{widetext}
\begin{equation}\label{eq:errorPLIF}
 \mathcal{L}_\text{data}=\frac{1}{2\left|\bfX^{\text{PLIF-PIV}}\right|} \sum_{\bfx\in\bfX^{\text{PLIF-PIV}}}	\Biggl(\left|\rho_{\mathcal{N}}(\bfx) - \rho_\bfx\right|^{2}+	\left|u_{\mathcal{N}}(\bfx) - u_\bfx\right|^{2} + 	\left|v_{\mathcal{N}}(\bfx) - v_\bfx\right|^{2}  \Biggr).
\end{equation}
\end{widetext}

Intuitively, $ \mathcal{L}=\mathcal{L}_\text{data}$ would be a poor choice for the flow reconstruction problem~:
it does not take into account variations in the $z$-direction and do not evaluate the model prediction for $w_{\mathcal{N}}$ and $p_{\mathcal{N}}$. Eventually, after solving \eqref{eq:optprob} the resulting model $\mathcal{N}$ would only fit $\rho$, $u$ and $v$ in the observational plane and would not generalize in the rest of the domain.

To address this issue, PINNs propose an enrichment of the objective function \eqref{eq:errorPLIF} based on the physical laws knowledge. This consists in minimizing the residuals $e_1$-$e_5$ of the governing equations, namely the LHS of  \eqref{eq:momx}-\eqref{eq:cont}, on a collection of points $\bfX^\text{res}$ that are referred to as collocation points. Whilst collocation points can be chosen arbitrarily \cite{raissi2018hidden}, we consider $\bfX^\text{res}=\mathcal{M}$, the corresponding loss term denoted by $\mathcal{L}_\text{res}$ reads~:
\begin{equation}\label{eq:errorPDE}
\mathcal{L}_\text{res}=\frac{1}{2\left|\bfX^\text{res}\right|} \sum_{\bfx\in\bfX^\text{res}} \sum_{q=1}^5	\left|e_q(\bfx)\right|^{2}.
\end{equation}

The automatic differentiation~\cite{baydin2018automatic} (AD) is leveraged to compute the time and space derivatives in $e_1$-$e_5$. AD is a direct application of the chain rule to neural networks that are composite function \eqref{eq:comp} , it allows computation of derivatives at machine precision and is computationally efficient. A special case of AD known as the backpropagation \cite{rumelhart1986learning}  has been notably the cornerstone of gradient-based learning algorithms for solving \eqref{eq:optprob}, which are implemented in modern deep learning libraries such as Tensorflow and Pytorch.

The complete objective function considered in this work is expressed as~:
\begin{equation}\label{eq:lossPINN}
\mathcal{L} = \mathcal{L}_\text{res}+\omega_\text{data} \mathcal{L}_\text{data}  +\omega_\text{bc}\mathcal{L}_\text{bc},
\end{equation}
where $\omega_\text{data}$ and $\omega_\text{bc}$ are a weighting coefficients and $\mathcal{L}_\text{BC}$ is an additional loss term that enforces the boundary conditions. For inverse problems \cite{raissi2020science,espresso}, the term $\mathcal{L}_\text{BC}$ is usually not included in the objective function. However in this work, the region of interest $\Omega_{\text{obs}}$ contains walls ($z=0$ and $y=0$) and boundary conditions are considered as additional data.

To conclude this section we define the term $\mathcal{L}_\text{data}$ in the \emph{LAT case} for which the training dataset $\mathcal{D}^{\text{LAT}}$ is composed of spanwise-averaged density measurements. The loss term is expressed as~:

\begin{equation}\label{eq:errorLAT}
 \mathcal{L}_\text{data}=\frac{1}{2\left|\bfXX^{\text{LAT}}\right|} \sum_{\bfxx\in\bfXX^{\text{LAT}}}\left|\overline{\rho}_{\mathcal{N}}(\bfxx) -\overline{\rho}_{\bfxx} \right|^{2},
\end{equation}
where the average density $\overline{\rho}_{\mathcal{N}}(\bfxx)$ at the point $\bfxx\in\bfXX^{\text{LAT}}$ is computed with the trapezoidal rule~:
\begin{equation}\label{eq:traplat}
\overline{\rho}_{\mathcal{N}}(\bfxx)=\frac{L_z}{2(N_z-1)}\sum_{k=1}^{N_z-1}\biggl(\rho_{\mathcal{N}}(\bfxx,z_{k})+\rho_{\mathcal{N}}(\bfxx,z_{k+1})\biggr).
\end{equation}
 Whilst the expression \eqref{eq:errorPLIF} is mathematically similar to \eqref{eq:errorLAT}, the nature of the data significantly differs. On one hand, $\rho_\bfx$ contains \emph{a priori} information with respect to the $z$ coordinate ($\bfx\in\mathbb{R}^4$), and on the other hand this information is absent in $\overline{\rho}_{\bfxx}$  ($\bfxx\in\mathbb{R}^3$) but encoded \emph{a posteriori} in formula \eqref{eq:traplat}. In Ref.~\onlinecite{mao2020physics}, the authors employed the trapezoidal rule to enforce the global conservation of mass and momentum for solving Euler equation with PINNs. 
 
 For the remaining test cases described in Section~\ref{sec:testcase}, the loss term $\mathcal{L}_\text{data}$ expression can readily be deduced from \eqref{eq:errorPLIF} and \eqref{eq:errorLAT}.

\begin{figure*}  
    \includegraphics[scale = 0.4]{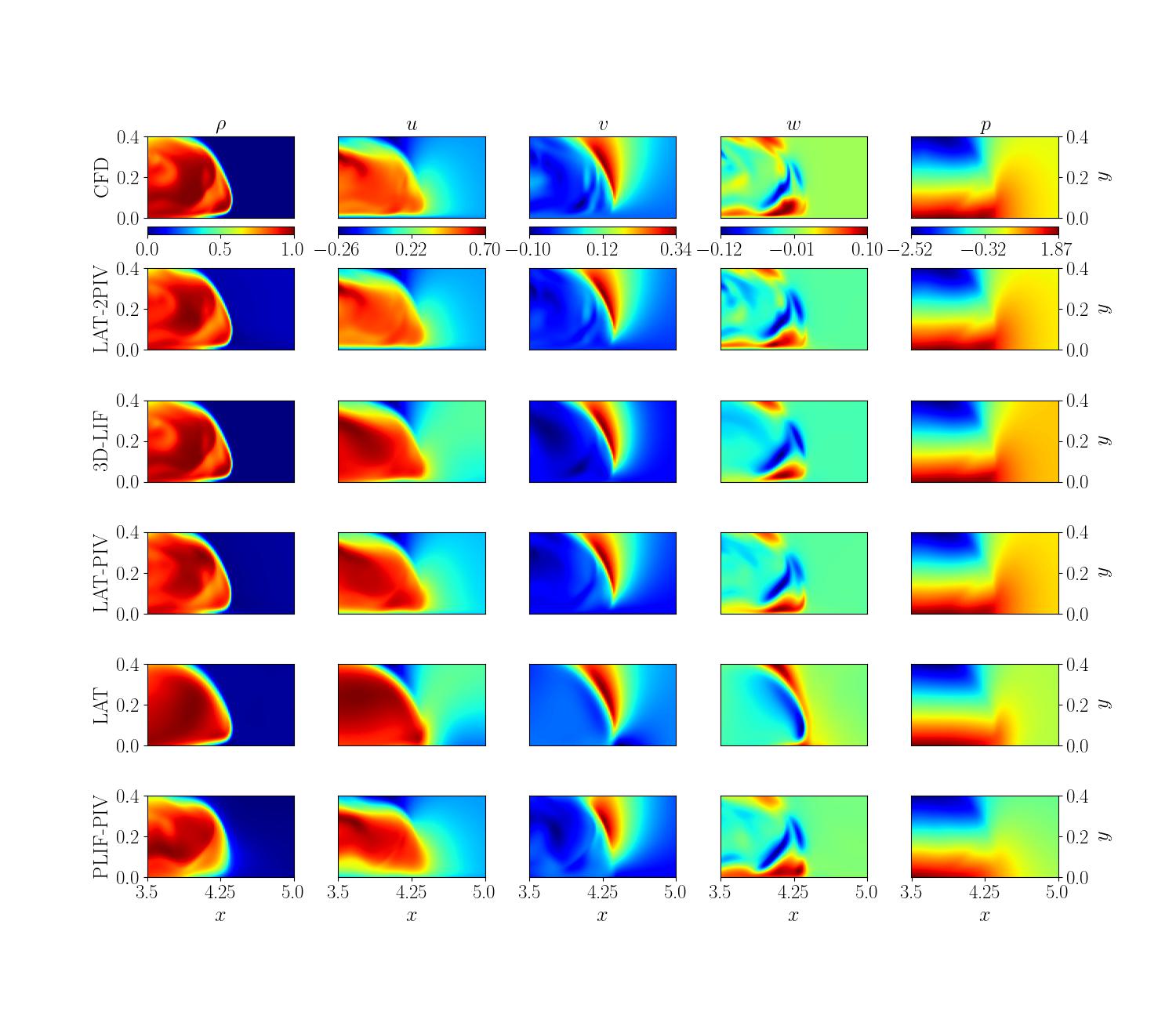}
    \caption{Isovalues of the reconstructed fields $\rho$, $u$, $v$, $w$ and $p$ at $z=0.154$ and $t=8.5$.
     }
    \label{fig:snapZ1}
\end{figure*}

\subsection{\label{hyperparam}Training PINN}
Before starting the training, several hyperparameters have to be chosen. These parameters encompass variables that determine the network structure and variables involved for solving the optimization problem~\eqref{eq:optprob}. A coarse grid search has been conducted over the hyperparameters space, which is not reported here. We give the retained values that achieve a good compromise between the model accuracy and the CPU time.

The neural network $\mathcal{N}$ is composed of $8$ hidden layers with $250$ neurons and the Swish activation function~\cite{swishramachandran2017searching} is used for each hidden layer, while no activation function is applied on the output layer. The hyperbolic tangent activation has been considered as well but it leads to slightly less accurate results. 

Among the hyperparameters introduced specifically by the PINN paradigm, the weight coefficients $\omega_\text{data}$ and $\omega_\text{bc}$ play an important role with respect to predictions accuracy. As in Ref.~\onlinecite{espresso}, constant values are considered and in the following we have $\omega_\text{data}=500$ and $\omega_\text{bc}=1$ unless otherwise specified. 
\begin{figure*}[ht!]
\centering 
   \includegraphics[scale = 0.35]{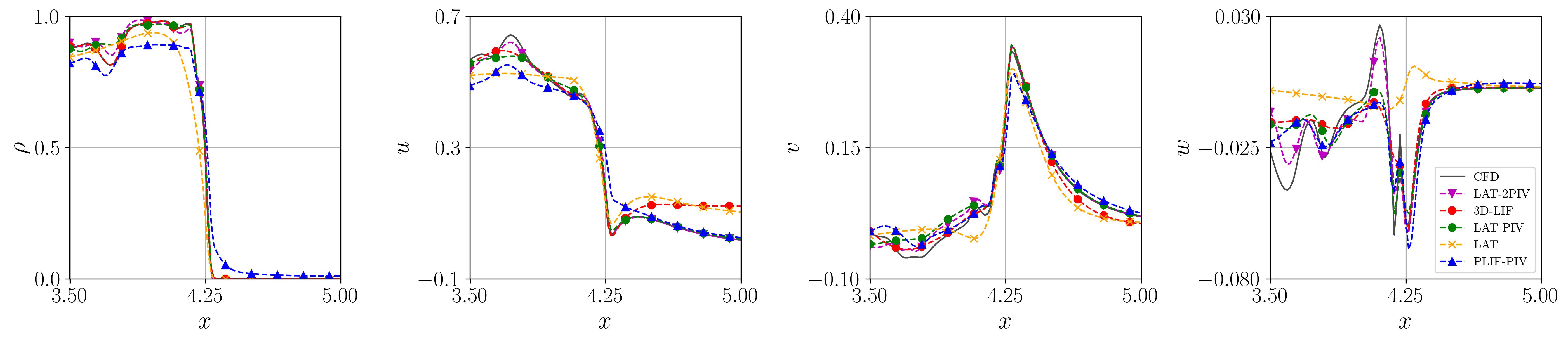} 
    \includegraphics[scale = 0.35]{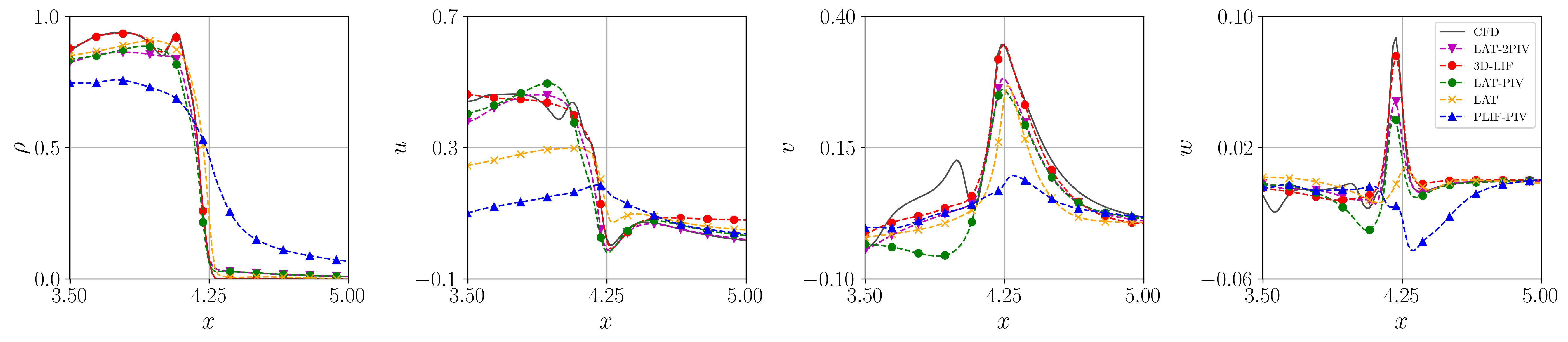}    
    \caption{Profiles of the reconstructed fields along the $x$-direction at $y=0.24$, $t=8.5$ and $z=0.154$ (top), $z=0.07$ (bottom).}
    \label{fig:prof_y}
\end{figure*}
The stochastic Adam optimizer~\cite{kingma2014adam} is employed for solving \eqref{eq:optprob}~: Glorot normal initializer~\cite{glorot2010understanding} is employed to initialize the biases and the weights that are computed iteratively with a gradient update (i.e., $\Theta_{n+1} = \Theta_{n} - \eta \nabla_{\mathcal{B}} \mathcal{L}$) on a subset $\mathcal{B}$, called a batch, of a given training dataset $\mathcal{D}$. One training round over the dataset is called an epoch. The learning rate is initially $\eta=5 \times 10^{-4}$ and as the training progresses, is lowered  by mean of an exponential decay scheduler. The final learning rate is $\eta=10^{-5}$ when training is stopped after $300$ epochs. We use a batch-size $|\mathcal{B}|$ of $4096$ and the network is trained with an Intel Xeon W2295 4.6Ghz CPU and a NVIDIA RTX A6000 GPU.

\section{\label{results}Numerical results}

The flow fields inferred by PINN from the datasets described in Section~\ref{sec:testcase} are now compared in terms of accuracy.
 After training, each model is evaluated on a  spatio-temporal Cartesian grid $\mathcal{M}'$ which contains three times as many points in each space direction and five times as many points in time as  $\mathcal{M}$. The fine mesh is used for visualization purpose and to evaluate the accuracy of the inferred flow fields, which is achieved with the relative $L_2$ error~:
\begin{equation}\label{eq:L2gene}
\epsilon_V =\frac{100}{\sup_{\bfx \in \mathcal{M}'} \left| V_\bfx \right|}\sqrt{\frac{\sum_{\bfx\in\bfX}	\left|V_{\mathcal{N}}(\bfx) - V_\bfx\right|^{2}}{\left|\bfX\right|}},
\end{equation}
where $V \in (\rho,u,v,w,p)$ and $\bfX\subseteq\times\mathcal{M}'$ defines the domain where the error is computed. For instance, if $\bfX=\mathcal{M}'$ then $\epsilon_V$ defines a global metric on the whole space and time domain. We choose $\sup_{\bfx \in \mathcal{M}} \left| V_\bfx \right|$ as reference value to avoid the division near to zero issue reported in Ref.~\onlinecite{cai2022physics}. 

\noindent Our preliminary tests have shown that including BCs on the bottom wall $y=0$ negatively impacts on the overall accuracy and thus have been discarded from $\mathcal{L}_\text{bc}$. On the contrary, integrating BCs on the wall $z=0$ improves significantly the accuracy for the cases \emph{LAT}, \emph{PLIF-PIV} and \emph{LAT-PIV} but has no influence on the other studied cases.

% \begin{figure*}  
%     \includegraphics[scale = 0.48]{Images_LowRes/z_39_sur_18_0.154_t_25_sur_50.jpg}
%     \caption{Isovalues of the reconstructed fields $\rho$, $u$, $v$, $w$ and $p$ at $z=0.154$ and $t=8.5$.
%      }
%     \label{fig:snapZ1}
% \end{figure*}

%\clearpage

\subsection{Inference accuracy}\label{sec:infres}
% For clarity, the density, velocity and pressure inferred from the observed data provided in the \emph{X} case will be referred to as \emph{X/PINN} hereafter e.g., \emph{LAT/PINN}.
The overall reconstruction error is reported in Table \ref{table:overall} and for all cases, the error is lower than $18\%$. In Fig.~\ref{fig:snapZ1}, we show the reconstructed fields in the vicinity of the symmetry plane ($z=L_z/2$). It is clear that the PINN approach allows to recover the main flow structure and provides a description of the pressure $p$ and the velocity $w$ while being absent from the training datasets. 
We stress out that for the case \emph{3D-LIF}, the density field is simply regressed thus the corresponding error is very low ($\simeq0.5\%$) and the density field is indistinguishable from the reference solution. 
The pressure has proven to be inferred accurately for all cases, with an error less than $10\%$, and varies little in the $z$-direction thus we focus the rest of the discussion on the density and velocity fields.

\begin{table}[H]
\caption{\label{table:overall}Relative $L_2$-norm errors of the fields reconstructed by PINN computed on the spatio-temporal Cartesian grid $\mathcal{M}'$.}
    \centering 
     \begin{tabular}{ |p{1.6cm}|c c c c c|  }
        %\hline
        %\multicolumn{7}{|c|}{ Global relative error $\epsilon$ (\%)} \\
         \hline
        Case & $\epsilon_\rho$ & $\epsilon_u$  & $\epsilon_v$  & $\epsilon_w$ & $\epsilon_p$  \\
         \hline
         LAT-2PIV &6.83\% & 3.31\% & 5.63\% & 3.91\% & 2.73\%  \\
         3D-LIF  & 0.51\% & 7.96\% & 5.16\% & 4.94\% & 3.85\%  \\
         LAT-PIV & 7.98\% & 5.65\% & 8.25\% & 6.10\% & 2.17\% \\
        LAT & 9.80\% & 13.41\% & 11.38\% & 7.59\% & 7.77\% \\
         PLIF-PIV & 17.11\% & 15.95\%  &  13.38\% & 7.37\% & 11.75\% \\
        \hline
    \end{tabular}   
    
\end{table}

The evolution of the error as a function of the time and space coordinates is given in Fig.~\ref{fig:distriberror} of Appendix~\ref{app:distrierror}. In general, the observed error depends little on the $x$ and $t$ coordinates but depends significantly on the $y$ and $z$ coordinates. Indeed, we observe that the error is concentrated in the near-wall zone $y=0$ and $z=0$ which correspond to zones of steep gradients. In particular, for \emph{PLIF-PIV} the reconstruction error can reach up to $20\%$ which results in unrealistic predictions, as shown by the reconstructed fields at $z=0.034$ in Fig.~\ref{fig:snapZ0} of Appendix \ref{app:isowall}.

Surprisingly for case \emph{LAT}, which corresponds to the smallest dataset, we are able to infer the main flow features and the errors are still acceptable. This illustrates the potential of spatially-averaged data in the context of flow reconstruction by PINNs. Indeed combining \emph{LAT} observations with one and two additional 2D-2C PIV measurements improves dramatically the accuracy of the inference : the results of the \emph{LAT-PIV} case outperform the \emph{PLIF-PIV} ones and for the case \emph{LAT-2PIV} the accuracy of the reconstruction is equivalent or better than that obtained for the case \emph{3D-LIF},  with notably an error on $u$-velocity twice as low, see Table~\ref{table:overall}. 

\begin{figure*}[ht!]
    \centering 
    \includegraphics[scale = 0.43]{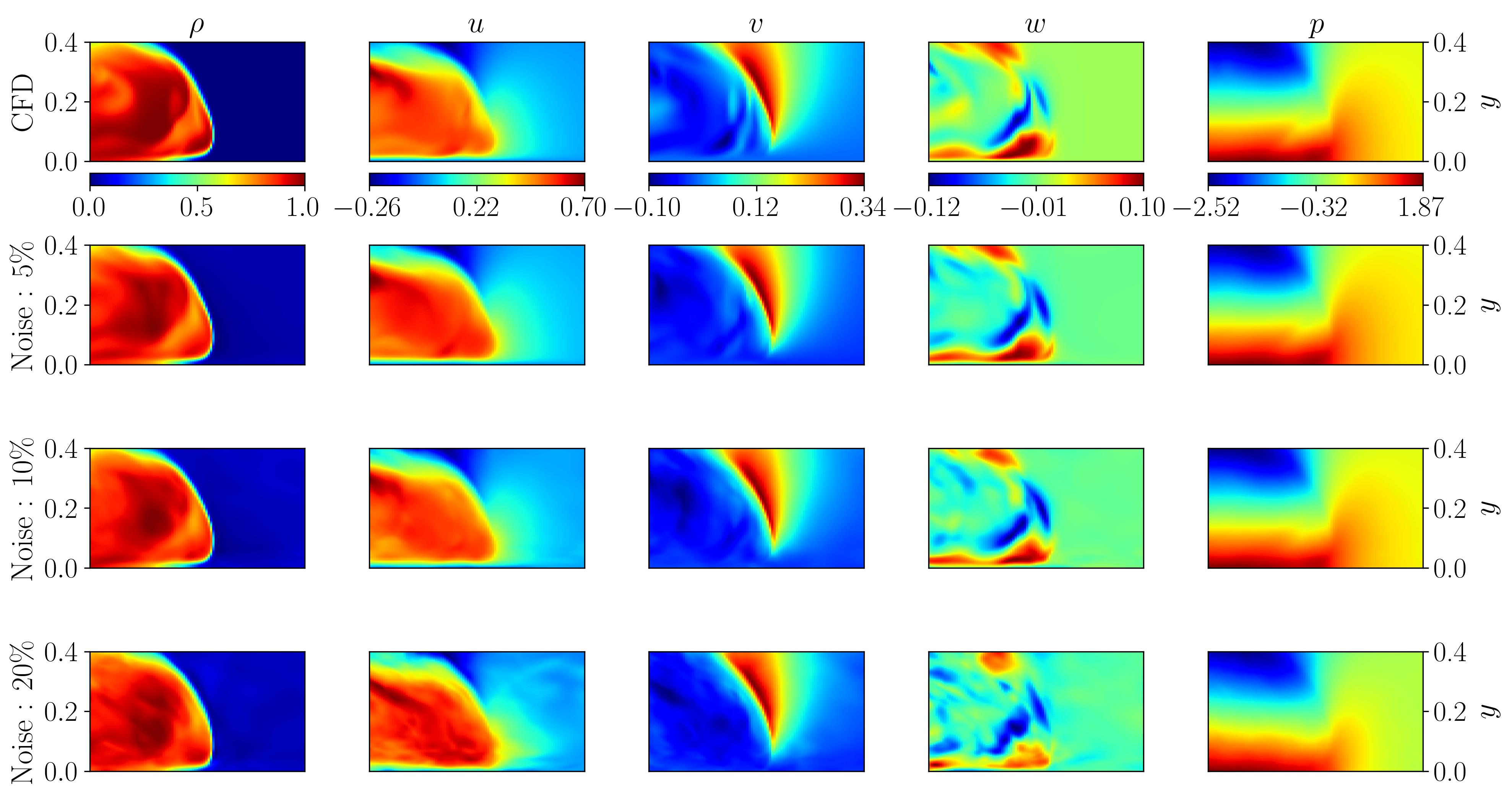}
    \caption{Isovalues of the reconstructed fields from noisy training data at $z=0.154$ and $t=8.5$. }
    \label{fig:isonoisy}
\end{figure*}

In order to compare further the inference results, we present in Fig.~\ref{fig:prof_y} the flow profiles along the $x$-direction at $z=0.154$ (upper plot) and in the near wall region $z=0.07$ (lower plot). We observe a good agreement with the reference results at $z=0.154$ for all cases inside the gravity current. For $x>4.3$, the $u$-velocity is overpredicted for cases \emph{LAT} and  \emph{3D-LIF}, for which only density observations are available. Since in this zone the density does not vary significantly ($\rho\simeq0$), we suspect that the reconstruction setups suffer from a lack of relevant observational data in this area, leading to poor predictions. For the case \emph{LAT-2PIV}, an excellent agreement with the reference results is observed for all velocity components. In particular, despite the low magnitude of the $w$-velocity, all variations are accurately captured.

Near the wall at $z=0.07$, the quality of the predictions deteriorates according to the evolution of the error given by Fig.~\ref{fig:distriberror} of Appendix~\ref{app:distrierror} and only the reconstruction setups \emph{LAT-PIV}, \emph{LAT-2PIV} and \emph{3D-LIF} allow to obtain an overall good agreement with the reference results. With these models we are able to predict accurately the density and the $u$-velocity fields. For the velocity components $v$ and $w$ on the other hand, only the maximum value at $x\simeq4.25$ is predicted satisfactorily and inside the gravity current none of the models allows to capture the secondary velocity peaks. If the origin of this loss of accuracy is not formally established, it is reasonable to think that the presence of steep gradients in vicinity of the wall combined with the low magnitude of the velocities $v$ and $w$  have a negative impact on the quality of the PINN models. In the field of CFD, loss of accuracy in boundary layers is a well-known problem for which a simple remedy consists in increasing the mesh density, see for instance~\cite{wesseling2009principles}. By analogy, dynamically allocating collocation points $\bfX^\text{res}$ during training by densifying them, for example, where the residuals $e_{1,4}$ are highest could improve the accuracy of PINNs models. Such algorithm has been recently proposed~\cite{wu2023comprehensive} and its implementation is left for a future work.

\subsection{\label{sec:noise}Influence of noisy data}

We investigate the robustness of the reconstruction algorithm by considering a training dataset that simulates noisy experimental data. To this purpose, the training dataset $\mathcal{D}^{\text{LAT-2PIV}}$ is corrupted with Gaussian noise with a noise level ranging from low to moderate, namely $5\%$, $10\%$ and $20\%$. The corresponding training data are represented in Figure~\ref{fig:noisydata} of Appendix~\ref{app:noise}.
\begin{table}[H]
\caption{\label{table:noise}Relative $L_2$-norm errors of the fields reconstructed by PINN in LAT-2PIV case, computed on the spatio-temporal Cartesian grid $\mathcal{M}'$ for different level of noise.}
    \centering 
     \begin{tabular}{ |p{1.9cm}|c c c c c|  }
        %\hline
        %\multicolumn{7}{|c|}{ Global relative error $\epsilon$ (\%)} \\
         \hline
        Noise level & $\epsilon_\rho$ & $\epsilon_u$  & $\epsilon_v$  & $\epsilon_w$ & $\epsilon_p$  \\
         \hline
         0 \% & 6.83\% & 3.31\% & 5.63\% & 3.91\% & 2.73\% \\
         5 \% & 8.00 \% & 4.41 \% & 6.07 \% & 4.40\% & 2.84\%  \\
         10 \%  & 8.41 \% & 5.53\% & 6.46\% & 4.78\% & 3.66\%  \\
         20 \% & 9.57 \% & 7.63\% & 8.12\% & 6.13\% & 7.03\% \\
        \hline
    \end{tabular}       
\end{table} 
As reported in Table~\ref{table:noise}, the reconstruction error is relatively low and a maximal value of approximately $10\%$ is observed for the density field for a noise level of $20\%$. The reconstruction error increases slowly with the noise level considered to reach a value of $10\%$ for a noise level of $20\%$. As shown in Figure~\ref{fig:isonoisy}, even at moderately high noise levels the PINN models are able to infer a very satisfactory level of detail in the gravity current. This confirms the high robustness to noisy data of the reconstruction procedure by PINN, which has been also reported in previous studies~\cite{raissi2020science, raissi2018hidden}.

\section{\label{conclusion}Conclusion}
This work is an application of the PINN method for flow reconstruction in the context of gravity currents. Based on a numerical experiment on the lock-exchange configuration, we benchmark several reconstruction setups that mimic experimental measurement techniques such like PIV, LIF and LAT. The accuracy of the flow reconstruction highly depends on the nature and the size of the dataset used for the  PINN training. When volumetric measurements of the density are available (\emph{3D-LIF} case) PINNs can infer the pressure and velocity fields very accurately and up to near wall regions. However the experimental cost of a 3D-LIF system is prohibitive and experimentalists more often employ PLIF systems combined with planar PIV measurements for studying gravity currents, which corresponds to the \emph{PLIF-PIV} case in this study. According to our results, this experimental setup seems to be not very suitable for PINN flow reconstruction: the accuracy of the inference decreases significantly as one moves away from the observation plane, leading to unrealistic predictions near the side wall.
Surprisingly, more accurate predictions are obtained by training the PINNs only from spanwise-averaged observations of the density by LAT, which can be integrated in the PINN loss function by employing the trapezoidal rule. Easy to implement experimentally, the LAT turns out to be an excellent building block for designing an experimental apparatus augmented with PINNs. Towards this direction we consider the \emph{LAT-2PIV} case for which LAT observations and two PIV planes are available, which allows a reconstruction as accurate as in the \emph{3D-LIF} case for a fraction of the experimental cost. Our results suggest that PINNs are an efficient and promising tool for flow reconstruction and we think that a new class of PINNs enhanced metrological devices may emerge in the near future.

\section*{\label{sec:acknowledgements}Acknowledgements}

We acknowledge support from  "PowderReg" project, funded by the European programme Interreg VA GR within the priority axis 4 "Strengthen the competitiveness and the attractiveness of the Grande Région / Gro$\beta$region. A part of this study is conducted in the framework of the Agence Nationale de la Recherche/DFG (Project "Pastflow" ANR- 19-CE08-0030-01).
\section*{Data Availability Statement}
The data that support the findings of this study are available from the corresponding author upon reasonable request.  
\clearpage

\begin{widetext}
\appendix
\section{\label{app:distrierror}Error distribution}
Figure~\ref{fig:distriberror} gives the evolution of the error as a function of the time and space coordinates.
\begin{figure*}[h!]
    \centering 
    \includegraphics[scale = 0.33]{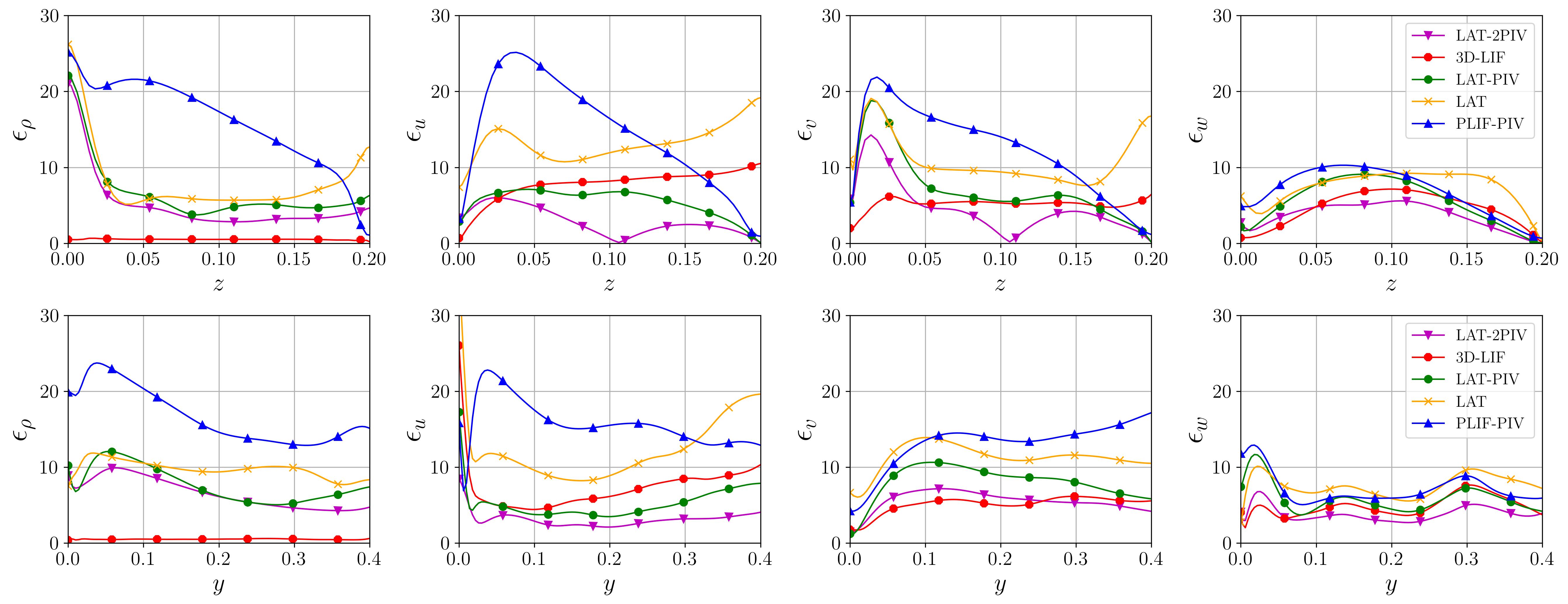}
    \includegraphics[scale = 0.33]{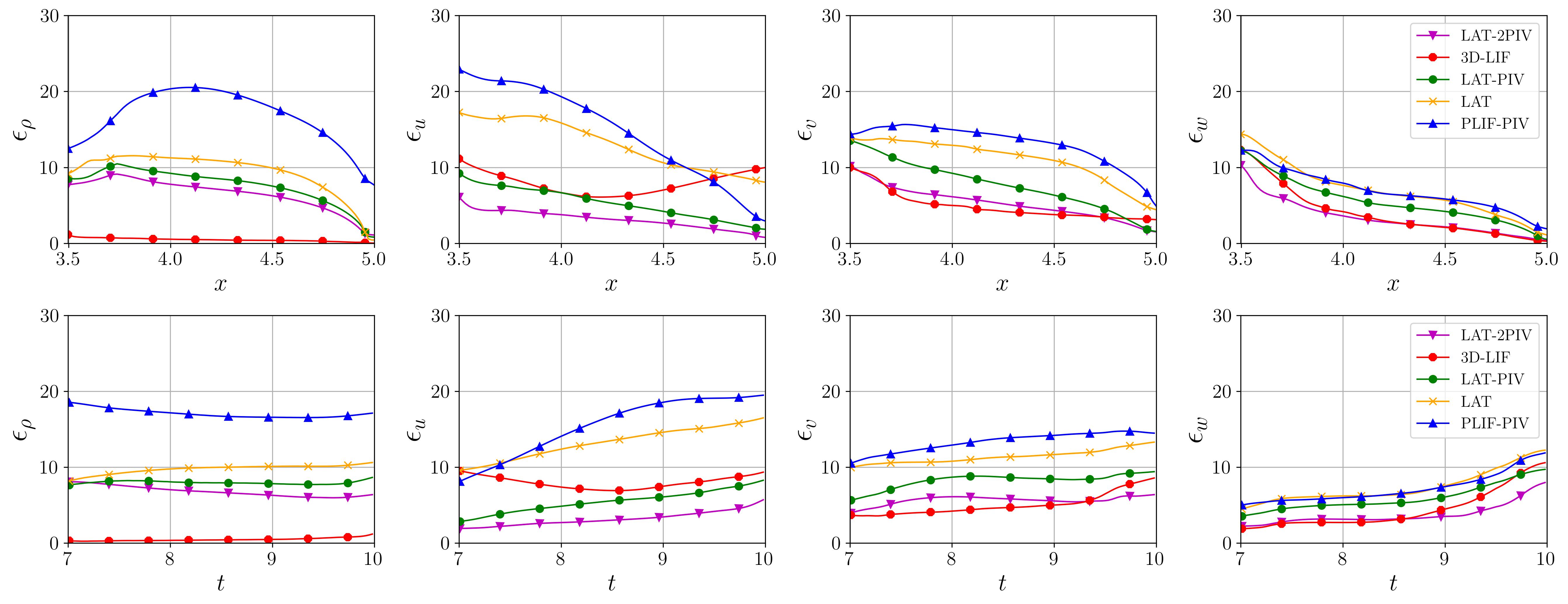}
    \caption{Evolution of the relative $L_2$-norm errors as a function of space and time coordinates for the different reconstruction setups. }
    \label{fig:distriberror}
\end{figure*}

\section{\label{app:isowall}Inference results near the wall}
We show the reconstructed fields in the vicinity of the wall $z=0$ in Fig.~\ref{fig:snapZ0}.
\begin{figure*}[h!]
    \includegraphics[scale = 0.4]{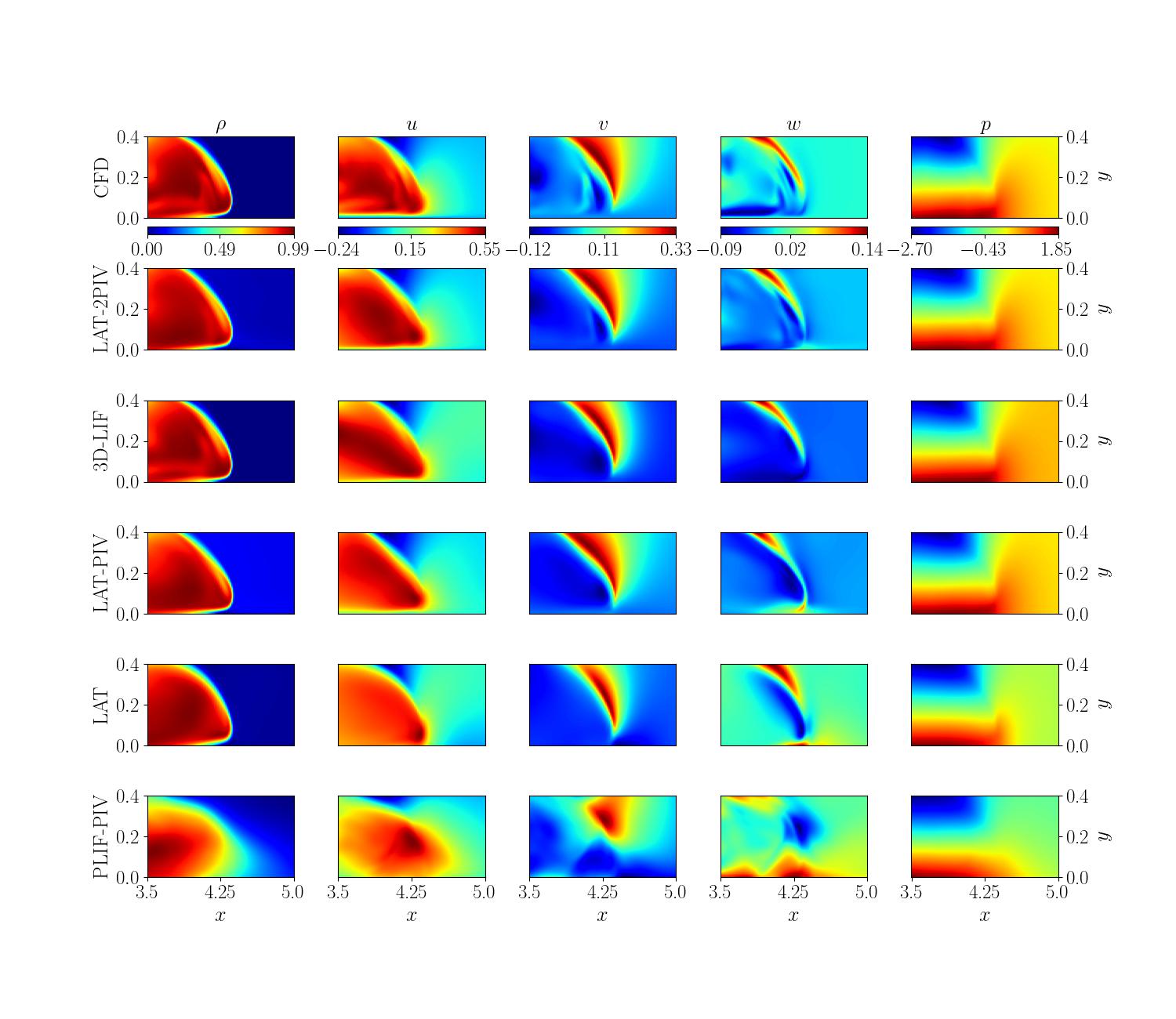}
    \caption{Comparison of $\rho$, $u$, $v$, $w$ and $p$ inferred by PINN from different datasets with the reference results at $z=0.034$ and $t=8.5$.
     }
    \label{fig:snapZ0}
\end{figure*}

\newpage

\section{\label{app:noise}Noisy training data}
Figure~\ref{fig:noisydata} shows the training data corrupted using synthetic additive Gaussian noise.

\begin{figure*}[ht!]
    \centering 
    \includegraphics[scale = 0.33]{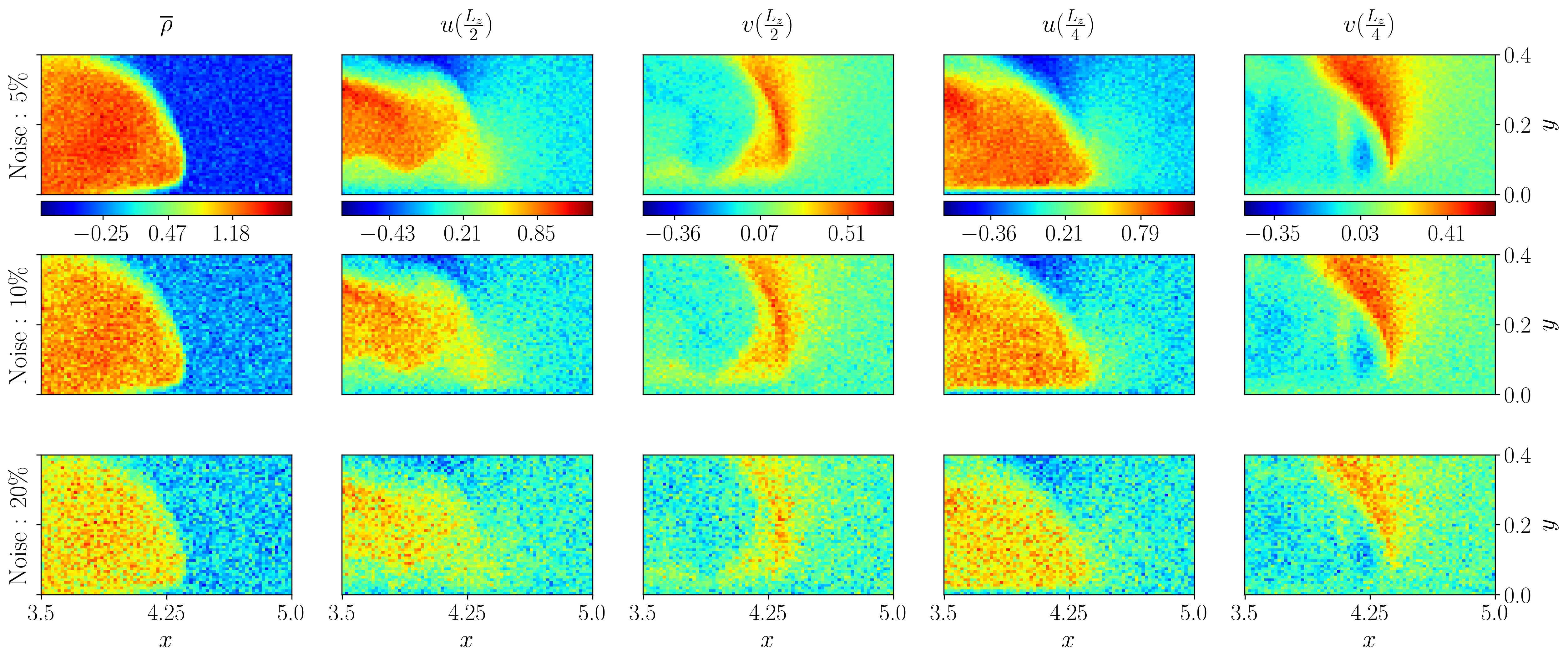}
    \caption{Training data with Gaussian noise for the \emph{LAT-2PIV} case at $t=8.5$. }
    \label{fig:noisydata}
\end{figure*}
\end{widetext}
\nocite{*}

\clearpage

\bibliography{biblio}% Produces the bibliography via BibTeX.

\end{document}